\documentclass[10pt]{article}
\usepackage{graphics,epsfig}

\font\smallroman=cmr9

\begin{document}
\title{{\bf Understanding Quantum Mechanics Through the\\Complementary Descriptions Approach}}
\author{{\sc Christian de Ronde}}
\date{}
\maketitle \centerline {Center Leo Apostel (CLEA) and Foundations
of the Exact Sciences (FUND)} \centerline {Faculty of Science,
Brussels Free University} \centerline {Krijgskundestraat 33, 1160
Brussels, Belgium.} \centerline {cderonde@vub.ac.be}

\begin{abstract}
\noindent Niels Bohr introduced the concept of complementarity in
order to give a general account of quantum mechanics, however he
stressed that the idea of complementarity is related to the
general difficulty in the formation of human ideas, inherent in
the distinction between subject and object. Earlier, we have
introduced a development of the concept of complementarity which
constitutes a new approach to the interpretation of quantum
mechanics. We argue that this development allows a better
understanding of some of the paradigmatic interpretational
problems of quantum theory.

Within the scheme proposed by modal interpretations we analyze the
relation between holism and reductionism as well as the problems
proposed by Arntzenius and Clifton. We discuss the problem of
presupposing the concept of entity within the quantum formalism
and bring into stage the concept of faculty as a way to recover
the objective character of quantum mechanics.
\end{abstract}\bigskip
\noindent

This article is dedicated to Kalervo Laurikainen who introduced me
to the deep thought of Wolfgang Pauli. It is these ideas which
have guided this work.

\newtheorem{theo}{Theorem}[section]

\newtheorem{definition}[theo]{Definition}

\newtheorem{lem}[theo]{Lemma}

\newtheorem{prop}[theo]{Proposition}

\newtheorem{coro}[theo]{Corollary}

\newtheorem{exam}[theo]{Example}

\newtheorem{rema}[theo]{Remark}{\hspace*{4mm}}

\newtheorem{example}[theo]{Example}

\newcommand{\proof}{\noindent {\em Proof:\/}{\hspace*{4mm}}}

\newcommand{\qed}{\hfill$\Box$}

\newcommand{\ninv}{\mathord{\sim}} 

\tableofcontents

\newpage

\section{The Problem of Understanding Quantum Mechanics}

Quantum mechanics was born in one of the most turbulent periods in
the history of western philosophy. At the end on the XIX century
the classical conception of the world was threatened, critics
started to spread out from different disciplines: philosophy,
mathematics, literature, music, etc. In physics, right at the
beginning of the XX century, a new theory accomplished the
unthinkable and escaped the limits of classical physical reality.
Quantum mechanics raised from the depths of the Tartarus, a dark
land of thought where the lightnings of Zeus had never arrived.
Since then, and more than one century later, the problem to which
quantum mechanics confronts us remains untouched. The famous
statement of Richard Feynman (\cite{Feynman92}, p.129) gives an
idea of the state of affairs: \emph{``I think it is safe to say
that no one understands quantum mechanics."}

The epistemological constraints imposed by the way of interacting
with atoms opened the doors of a new world. The way in which
atomic phenomena was acquired was formally expressed by the
quantum postulate which, according to Bohr, expresses the most
important character of the atomic theory:

{\smallroman
\begin{quotation}
``The quantum theory is characterized by the acknowledgement of a
fundamental limitation in the classical physical ideas when
applied to atomic phenomena. [...] its essence may be expressed in
the so called quantum postulate, which attributes to any atomic
process an essential discontinuity, or rather individuality,
completely foreign to the classical theories and symbolized by
Planck's quantum of action." N. Bohr (quoted from \cite{WZ},
p.88)\end{quotation}}

Werner Heisenberg was able to extend this critic into a consistent
formalism which represented the new epistemological constrains
imposed by experiencing with atoms. As expressed by Bohr:

{\smallroman
\begin{quotation}
``As is known, the new development was connected in a fundamental
paper by Heisenberg, where he succeeded in \emph{emancipating
himself completely from the classical concept of motion} by
replacing from the very start the ordinary kinematical and
mechanical quantities by symbols which refer directly to the
individual processes demanded by the quantum postulate." N. Bohr
(quoted from \cite{WZ}, p.105, emphasis added)
\end{quotation}}

It was the critic to the classical concepts which guided
Heisenberg into the principle of indetermination of quantum
observables. As Heisenberg recalls in his autobiography \emph{Das
teil und der ganze} \cite{Heisenberg69}, it was Albert Einstein
himself who inspired him by expressing the idea that: \emph{``it
is only the theory which decides what can be observed."} In this
way the theory appears as the condition of possibility to access
certain phenomena. Quantum mechanics, escaping from the
presuppositions of classical physics, had determined a \emph{new
experience}, a\emph{ new physics}.

Even though a discipline such as physics is historically
constrained by practices, practices which pertain to a certain
epoch, we should never forget the origin and development which
have pictured this discipline. Physics, like occidental culture,
was born in Greece. At the beginning of the VI century B.C. in
Miletus, mythical and theological explanation changed into
rational reflection about Nature. Thales, Anaximander and
Anaximenes started a systematical investigation, a
\textit{histor\'ia}, of which they presented a
\textit{the\~{o}ria} of the origin of the world, its composition,
its order. The first philosophers were called \emph{physicists} as
they related themselves to \emph{physis}; they placed the
fundament of existence outside the laws of the city. In opposition
to the Sophists, who believed that man was the measure of all
things, physicists placed the fundament of existence in Nature, in
universal laws which governed the cosmos, independently of the
wishes and desires of the Gods \cite{Vernant62}. Philosophers
distinguished between \textit{doxa}, mere opinion, and
\textit{episteme}, true knowledge, also translated as
\emph{science}.\footnote{I wish to thank Joaquin Piriz for the
many discussions on early Greek philosophy and its relation to
science.}

Physics is concerned with the question of understanding Nature.
The questions which guide this search have both an ontological and
epistemological character. Physics is not just a discourse about
Nature, but rather a discipline which tries to express Nature. A
physicist, we believe, is someone who admires the world and
wonders about its possibilities and impossibilities, its structure
and meaning. To \emph{understand} in physics means to have a
picture of the mathematical formalism which expresses Nature, a
map between the mathematical scheme and the concepts which explain
a certain experience. However, the known reflection of Einstein to
his friend Besso regarding light quanta: ``Nowadays every Tom,
Dick and Harry thinks he knows it [the answer to the question:
`What are light quanta?'], but he is mistaken." is applicable to
quantum mechanics itself. We do not know what quantum mechanics is
talking about. Until we do not answer this ontological question we
won't be able to say we have understood quantum mechanics.

In this paper we would like to present the {\it complementary
descriptions approach} as already discussed in \cite{deRondeCDI}
and \cite{deRondeCDII}. We consider this approach a development as
well as an extension of our reading of Niels Bohr, Wolfgang Pauli
and Werner Heisenberg, a general framework and a philosophical
worldview in which we attempt to understand some of the
interpretational problems present in modern physics. The
complementary descriptions approach takes into account the
different conceptual schemes needed to account for the paradoxes
which hunt the quantum formalism; it also stresses the need of
bringing into stage new concepts, which can expose in a new light
the seemingly weird character of quantum mechanics.

\section{The Complementary Descriptions Approach}

In \cite{deRondeCDI} we developed the concept of
\emph{complementarity} in order to take into account not only {\it
complementary contexts} (``phenomena" in Bohr's terminology) but
also {\it complementary descriptions} such as the {\it classical
description}, which stresses the reductionistic character of the
being, and the {\it quantum description}, which stresses the
holistic character of the being.\footnote{For a review of the
concept of holism and its relation to quantum mechanics see for
example \cite{Healey99, Seevinck04} and references therein.} For
this purpose we will follow a middle path between ontology, i.e. a
certain form of the being; and epistemology, i.e. the theoretic
preconditions of the description which determine a certain access
to the being. Ontology is an expression of the {\it being}, it is
not {\it the being} itself but its expression, thus, always of a
specific form, a restricted mode of the being. The main idea is to
take into account the distinction of different complementary
descriptions which refer to one and the same reality. Different
conceptual schemes (descriptions) define different ontologies
through their relation to experimental observation. The being
arises from the \emph{difference} between descriptions, just as
the one expresses itself through the many.\footnote{I am specially
grateful to F. Gallego for discussing this important point with
me.} Within the position taken by the complementary descriptions
approach different descriptions are able to create different
experiences. We want to stress the importance of knowing from
which description one is talking from, because it might happen
that the same sentence can have only significance in a certain
description while it is meaningless in a different one
(\cite{deRonde03}, section 9.3). We will argue that only together,
different complementary descriptions allow a resolution of the
paradoxes which hunt the interpretation of quantum mechanics.

\subsection{Descriptions, Perspectives, Contexts and Properties}

We would like now to give some definitions already stated in
\cite{deRondeCDI, deRondeCDII} in order to go further into the
discussion. Firstly, {\bf descriptions} are a general framework in
which concepts relate, they express the precondition to access a
certain expression of reality. A description is defined by a
specific set of concepts, this definition precludes, at a later
stage, the possibility of applying incompatible concepts to
account for the same phenomena. Quantum mechanics is a description
as well as classical mechanics or relativity theory are, each of
them relating concepts which are not necessarily compatible with
the ones present in a different description.\footnote{In this
sense my approach goes together with Heisenberg's conception of
{\it closed theories} as a relation of tight interconnected
concepts, definitions and laws whereby a large field of phenomena
can be described \cite{Bokulich04}.} Contrary to N. Bohr's idea,
that: {\it ``it would be a misconception to believe that the
difficulties of the atomic theory may be evaded by eventually
replacing the concepts of classical physics by new conceptual
forms"}\footnote{Quoted from (\cite{WZ}, p.7).} our approach
accompanies the line of thought of W. Pauli, who stressed the
importance of the development of more general thought-forms:

{\smallroman
\begin{quotation}
``If in spite of the logical closure and mathematical elegance of
quantum mechanics there is one part of some physicists a certain
regressive hope that the epistemological situation we have
sketched may turn out not to be final, this is in my opinion due
to the strength of traditional thought forms embraced in the
designation `ontology' or `realism'. Even those physicists who do
not reckon themselves entirely as `sensualists' or `empiricists'
must ask themselves the question, which it is possible to ask on
account of the postulational character of these traditional
thought-forms, and is unavoidable on account of the existence of
quantum mechanics -- namely the question whether these
thought-forms are necessary condition that physics should be
possible at all, or wether other, more general thought-forms can
be set up in opposition to them. The analysis of the theoretical
foundations of wave or quantum mechanics has shown that the second
alternative is the correct one." W. Pauli (\cite{Pauli94},
p.47)\end{quotation}}

In the orthodox formulation of quantum mechanics the wave function
is expressed by an abstract mathematical form. Its representation
can be expressed through the choice of a determined basis $B$. The
non-represented wave function is a {\bf perspective}. A
perspective expresses the {\it potentiality} of an action which
makes possible the choice of a definite context, it is the
condition of possibility for a definite representation to take
place; it deals with the choice between mutually
\emph{incompatible} contexts. The perspective cannot be written,
it shows itself through the different representations, each of
which is a part but not the whole. The importance of defining this
level of description is related to the structure of the quantum
formalism which, contrary to classical mechanics, is essentially
holistic and thus, intrinsically contextual; i.e., it does not
allow for the simultaneous existence of mutually incompatible
contexts. The {\bf context} is a {\it definite representation} of
the perspective, it depends and configures in relation to the
concepts which are used in the description.\footnote{The
distinction between perspective and context was introduced in
\cite{deRonde03} in order to distinguish between the different
modes of existence of the properties in the modal interpretation.
Following van Fraassen's distinction between dynamical and value
state, we have distinguished between holistic and reductionistic
contexts (see also Karakostas 2004 for a similar discussion).} The
different possible contexts can not be thought as encompassing a
whole of which they are but a part (see \cite{deRondeCDII},
section 1.2).

Let's try to understand these concepts through some examples. In
special relativity theory a context is given by a definite
inertial frame of reference. However, there is no need of defining
the perspective because the invariance principle, given by the
Lorentz transformations, allows us to think all these different
contexts as existing in actuality, as events which pertain to
physical reality. There is a way by which one can relate all the
events which actually exist in the same picture (even though their
relation is different form that of classical mechanics). In
quantum mechanics, on the other hand, a context is given by a
definite experimental set up; i.e. a complete set of commuting
observables (C.S.C.O.) which is defined equivalently by a quantum
wave function in a {\it definite representation/basis}. But
because of \emph{Heisenberg's principle of indetermination} there
is an intrinsic, ontological incompatibility between different
representations. The Kochen Specker theorem \cite{KS}, to which we
will return later, does not allow to think a property, which is
seen from different contexts, as existing in actuality. The
different contexts can not be thought in terms of possible views
of one and the same ``something". In linear classical mechanics
and special relativity theory, this problem does not arise because
one can relate contexts through the Galilean and Lorentz
transformations. One may say that in these theories one can reduce
all the different views to a \emph{single context},\footnote{Even
if we do not know the context we can think in terms of possible
contexts, in terms of ignorance.} and this is why the idea of
perspective becomes superfluous. The formal structure of classical
mechanics and relativity theory is reductionistic, and thus, part
and whole are equivalent.

It is only at the level of the context that one can speak of {\bf
properties}. Different set of properties arise in each
representation, which relate and are configured by the
\emph{logical principles} which govern the description. In the
case of classical mechanics properties relate via the principles
of classical (Aristotelian) logic; i.e. the principle of
existence, the principle of identity and the the principle of
non-contradiction (see section 4). The reductionistic character of
the structure arises from the choice of these ontological
principles, which at the same time, allows us to speak of
``something" which exists.\footnote{I wish to thank Karin Verelst
for the many discussions regarding this subject.} It is only
because one presupposes this structural configuration that one is
allowed to speak about {\it entities}. In quantum mechanics the
properties arising in each context do not follow classical
relationships but are determined by a different logic.
Heisenberg's principle of indetermination, Bohr's principle of
complementarity, Pauli's exclusion principle and the superposition
principle provide a structural relationship between
quantum-properties which cannot be subsumed into classical
thought.\footnote{For a detailed analysis and discussion of the
principles of indetermination, complementarity and superposition
as those which determine the fundamental logical structure of
quantum theory see \cite{LahtiBugajski80} and specially
\cite{Lahti80}.}

Different contexts cannot be thought as being faces of a single
perspective because the properties which arise in each context can
not be thought in classical terms, as existing in actuality. An
\emph{improper mixture} does not allow to think its elements as
actual or possible properties. A \emph{proper mixture}, on the
contrary, is ``something" which exists in actuality but of which
we are uncertain.\footnote{It should be noticed that a proper
mixture presupposes the existence of that of which it is talking
about, it deals with classical Kolmogorovian probability. As noted
by E. Schr\"{o}dinger (quoted from \cite{Bub97}, p.115):
``Probability surely has as its substance a statement as to
whether something {\it is} or {\it is not} the case --an uncertain
statement, to be sure. But nevertheless it has meaning only if one
is indeed convinced that the something in question quite
definitely {\it is} or {\it is not} the case. A probabilistic
assertion presupposes the full reality of its subject.''} The
procedure of measuring in quantum mechanics presupposes an
\emph{interpretational jump} going from an improper mixtures to a
proper mixture (something which Bernard D'Espagnat has proved
\cite{D'Espagnat76}, leads to inconsistencies). The problem is
that without this interpretational jump we could not make sense of
talking in terms of a classical apparatus. This is the path from a
{\it holistic context} with quantum holistic properties, with
superpositions, into a {\it reductionistic context} with classical
reductionistic properties and an ignorance interpretation (see
\cite{deRondeCDII}, sections 1.3 and 1.4).

When speaking of properties, one must recognize the discourse in
which they are embedded. In many discussions regarding the
interpretation of quantum mechanics one talks about quantum and
classical properties just like ``properties" without a proper
mention to its \emph{mode of being}, this lack of clarification
produces lots of pseudo-problems and misunderstandings which have
been discussed earlier (see \cite{deRondeCDII}, section 2).

The perspective has not been acknowledged in quantum mechanics due
to the wrong presuppositions which involve the characterization of
a quantum state as a vector in Hilbert space. In orthodox quantum
mechanics it is assumed that something like a ``vector" exists
(independently of the basis in which it is ``placed"). But in the
mathematical structure of quantum mechanics the basis plays an
{\it active} role, it constitutes the existence of the set of
properties which, at a later stage, determines that which will be
studied. ``This which will be studied", and can be best
characterized by an improper mixture, can not be subsumed into the
classical categories of an ``entity".\footnote{Note that one might
talk in a reductionistic context {\it as if} one would have an
entity. If we forget the procedure of successive cuts by which we
arrived at a holistic context with improper mixtures we might talk
{\it as if} this mixture is proper, and thus recover the logical
principles which allow us to talk about entities.} It is assumed
that the $\Psi$ contains all the different representations, that
it is in itself an identity, a unity, something which is able to
give account of the totality of the different representations
\cite{GdeR}. The ``same" vector however, cannot support the
existence of its different representations, precluding the
possibility of thinking of $\Psi$ in terms of something which
refers to an entity. As we will show in this paper, it is exactly
this idea which cannot be maintained in
quantum mechanics.\\

A brief outline of what we tried to explain until now can be
provided in the following scheme:\\

\begin{tabular}{|c|c|c|c|c|}
\hline
& \tiny{ \textbf{PERSPECTIVE}} & \textbf{ \tiny{HOLISTIC}} & \tiny{ \textbf{REDUCTIONISTIC}} & \tiny{ \textbf{MEASUREMENT}} \\
&  & \textbf{ \tiny{CONTEXT}} & \tiny{ \textbf{CONTEXT}} & \tiny{ \textbf{RESULT}} \\
\hline
\tiny{\textsl{MODE OF BEING}} & $?$ & $?$  & \footnotesize{ {\it possible/probable}} & \footnotesize{{\it actual}} \\
\hline
\tiny{\textsl{FORMAL EXPRESSION}} & $\Psi$ & $\psi_{B}$ & $\psi_{B}$ & $\alpha_{k}$, $|\alpha_{k}\rangle$ \\
\hline
\tiny{\textsl{THEORETICAL EXPRESSION}} & ? & \footnotesize{{\it improper mixture}} & \footnotesize{{\it proper mixture}} & \footnotesize{{\it single term}}\\
\hline
\tiny{\textsl{PROPERTY}} & -- & \footnotesize{{\it holisic/non-Boolean/}} & \footnotesize{{\it reductionistic/Boolean/}} & \footnotesize{{\it actual}} \\
 &  & \footnotesize{{\it superposition}} & \footnotesize{{\it ensemble}} &  \\
\hline
\tiny{\textsl{DESCRIPTION IN TERMS OF}} & ? & ? & \footnotesize{{\it possible entity}} & \footnotesize{{\it actual entity}} \\
\hline
\end{tabular}

\subsection{Convergence of Descriptions}

Complementarity goes with paradox, it allows us to stress the
limits of knowledge and, at the same time, it presents us with the
incommensurability of reality. Descriptions, perspectives,
contexts and concepts are then taken as complementary in this same
sense. With the complementary descriptions approach we try to find
a {\it middle path} between descriptions. The main difficulty of
this approach is to stand in between, to not be dragged by any
specific description, each of which should be regarded only as a
`partial description', and complementary to a different one.

There is a quite tacit assumption which goes against the ideas we
have been presenting, namely, the idea that science is converging
towards the ultimate truth, that our knowledge increases with
every paper that is published. We think the idea of what is
``understanding" has been severely damaged by a radical
positivistic attitude science has taken in the last centuries. One
should, as a scientist, wonder about the meaning of
``understanding".\footnote{See for example the very interesting
discussion between Heisenberg and Pauli regarding the concept of
`understanding' (chapter 6, \cite{Heisenberg69}).} This maybe the
core problem for many in reaching the concept of complementarity,
which tackles our tacit presuppositions in the traditional
positivist epistemological framework; i.e. that a theory provides
knowledge about an object if and only if it justifies making true
descriptive statements predicating properties of some substantial
entity \cite{Folse87}. We are interested in what \emph{is given}
in experience, something which should not be confused with what
one \emph{should expect} about experience \cite{GdeR}.

The idea of a convergent reality presupposes the idea that one can
reduce concepts of one theory to the next; i.e. that there is a
fundamental theory which can reach {\it the} fundamental `concepts
of Nature', the theory of everything. Richard Feynman is a
proponent of such view, in his BBC television lectures he argued:

{\smallroman
\begin{quotation}
``The age in which we live is the age in which we are discovering
the fundamental laws of nature, and that day will never come
again. It is very exciting, it is marvellous, but this excitement
will have to go." R. Feynman (quoted from \cite{Primas83}, p.347).
\end{quotation}}

Reductionism goes together with convergence. In this sense,
classical mechanics is worse than relativity theory, because, the
last is able to see the concepts of the first as a limit, and at
the same time it produces new insights. Contrary to this position,
it is quite clear that when one studies the problem in a deeper
way, one finds that such concepts are no limits, rather, they can
be found as approximations within certain very specific
conditions. But, when these conditions are extended to the general
frameworks from which the concepts acquire meaning,
incompatibilities and inconsistencies appear as much as in between
the classical and the quantum description. In other words, trying
to find a limit between quantum theory and classical mechanics is
to some extent equivalent in trying to find a limit between
physics and psychology. Although concepts like space and time can
be used in the physical framework as well as in the Freudian
theory of psychology, once we generalize the concepts to the
complete framework of either description we find out the concepts
generalize as well in both directions making impossible to retain
the consistency presented in the beginning. Quantum mechanics is
full of these type of mistakes which appear in most cases by using
concepts and symbols which are not part of the {\it quantum
description}.\footnote{A more detailed discussion has been
presented in \cite{deRondeCDII}.} Through mixing symbols and
concepts which pertain to different descriptions in many cases we
end up in weird paradoxes. In order to get closer to the mystery
one first needs to demystify and clarify the limits and the
correct usage of the different descriptions.\footnote{I am
specially grateful with Diederik Aerts for pointing out this road
sign to me.}

The idea that quantum mechanics is a fundamental theory of Nature
(as describing the fundamental blocks of reality from which
everything else can be derived), even the idea that there might
exist a true story about the world\footnote{See for example
\cite{Weinberg93}.} goes completely against the spirit of what we
are proposing here. This idea rests somehow on the presupposition
that science has reached (or might be able to reach) the {\it a
priori} conditions of human understanding itself.

{\smallroman
\begin{quotation}
``In many respects the present appears as a time of insecurity of
the fundamentals, of shaky foundations. Even the development of
the exact sciences has not entirely escaped this mood of
insecurity, as appears, for instance, in the phrases `crisis in
the foundations' in mathematics, or `revolution in our picture of
the universe' in physics. Indeed many concepts apparently derived
directly from intuitive forms borrowed from sense-perceptions,
formerly taken as matters of course or trivial or directly
obvious, appear to the modern physicist to be of limited
applicability. The modern physicist regards with scepticism
philosophical systems which, while imagining that they have
definitively recognized the {\it a priori} conditions of human
understanding itself, have in fact succeeded only in setting up
the {\it a priori} conditions of the systems of mathematics and
the exact sciences of a particular epoch." W. Pauli
(\cite{Pauli94}, p.95).
\end{quotation}}

\subsection{The Problem of (Re)Presentation}

Scientific realism is the position that theory construction aims
to give us a literally true story of what the world is like, and
that acceptance of a scientific theory involves the belief that it
is true. It is this idea of truth, as a closed enterprize, which
is responsible to great extent for the development of the `fabric
of science'. Reductionism allows a single truth, a single
description, it is hostile to every conception which is outside
its own limits. However, as Max Planck expressed in a very bright
way, science always finds its ways: ``A new scientific truth does
not triumph by convincing its opponents and making them see the
light, but rather because its opponents eventually die, and a new
generation grows up that is familiar with it." Quantum mechanics
presents us a new description, with new concepts which up to the
present have not been further developed.

The problem becomes more clear as we continue to analyze the
relation between the quantum and the classical
description.\footnote{See for example \cite{deRondeCDII}.} We
should note that, to have a (re)presentation of quantum mechanics
does not mean to place it within the classical domain (even though
one should explain in what sense a particular experience coincides
in the quantum and classical descriptions). The problem is not to
(re)present but ``to believe" in the (re)presentation, that the
(re)presentation is true, that having a representation allows us
to see through the veil of Maya. Even today the present approaches
to understand quantum mechanics take, explicitly or tacitly, our
classical conception of the world as the fundament. They believe
in the classical (re)presentation, and thus, try to explain
quantum mechanics with classical concepts. They try to find a {\it
limit} between quantum mechanics and classical mechanics.

One should be able to acknowledge the possibility of describing
the world from different (incompatible but complementary) view
points. Wolfgang Pauli imagined a future in which the quantum
conceptions would not be regarded as `weird.' This conceptual
jump, in the way we understand the world, has not yet taken place
with respect to quantum mechanics (nor relativity
theory\footnote{See Constantin Piron's reflection on this subject
in \cite{Piron}.}). We believe this is due to the lack of new
concepts in the quantum domain and to a `reductionistic
conception' which presupposes that `understanding' is reducible to
`classical understanding', to a single view point.\footnote{The
relation between quantum mechanics and classical concepts has been
investigated in \cite{deRondeCDII}.}

{\smallroman
\begin{quotation}
``[in quantum mechanics] there is no visualizable model
encompassing the whole structure [...] the demand that there
should be a visualizable model would be tantamount to demand that
classical physics should determine the conceptual tools of new
theories. This would deny the possibility of really new
fundamental theories, conceptually independent of classical
physics." D. Dieks (\cite{Dieks89}, p. 1417)\end{quotation}}

We might say that in order to provide a (re)presentation of
quantum theory we must go further into the development of new
concepts which escape the limits of classical thought and
accompany quantum mechanics to find its own interpretation. We
need to create a way of (re)presenting which is not classical. We
need to find new concepts which allow us to think the quantum
experience outside the bounds imposed by classical thought. The
problem of language, to which Bohr repeatedly referred, is an
expression of the collapse of our classical world. This is why, to
(re)present quantum mechanics means to confront what Pauli
considered the most important problem of our time:

{\smallroman
\begin{quotation}
``When the layman says ``reality" he usually thinks that he is
speaking about something which is self-evidently known; while to
me it appears to be specifically the most important and extremely
difficult task of our time to work on the elaboration of a new
idea of reality." W. Pauli (quoted from \cite{Laurikainen98},
p.193)
\end{quotation}}

\subsection{Development of New Concepts}

It must be clear at this point that we do not mean to take as a
standpoint, like Bohr did, the importance of classical (physical)
concepts in the definition of {\it experience}. Bohr relied in a
concept of complementarity which was a consistent explanation of
these {\it phenomena}. My approach is a development of the concept
of complementarity stressing the importance of {\it descriptions}
which make possible the preconditions of experimental observation
encouraging the creation and development of new concepts within
physics.

{\smallroman
\begin{quotation}
``In one of his lectures on the development of physics Max Planck
said: `In the history of science a new concept never springs up in
complete and final form as in the ancient Greek myth, Pallas
Athene sprang up from the head of Zeus.' The history of physics is
not only a sequence of experimental discoveries and observations,
followed by their mathematical description; it is also a history
of concepts. For an understanding of the phenomena the first
condition is the introduction of adequate concepts. Only with the
help of correct concepts can we really know what has been
observed." W. Heisenberg (\cite{Heisenberg73}, p.264)
\end{quotation}}

In physics, every new theory that has been developed, from
Aristotelian mechanics to general relativity, has been grounded in
new systems. {\it The physicist should be a creator of physical
concepts}. Concepts which, within a theory, make possible to grasp
certain character of Nature. This, however, should not be regarded
as some kind of solipsism, {\it it is not only the description
shaping reality but also reality hitting our descriptions.} It is
through this interaction, namely, our descriptions and the
experimental observation that a certain character of the being is
expressed. It is in this way that we can develop that which we
consider to be {\it reality}.

The main difference between quantum mechanics and the rest of the
theories created by man is that the quantum wave function
expresses {\it explicitly} a level in the description of Nature
which has been neglected from a mechanistic, entitative idea of a
clock-type-world. It presents us with the concept of {\it choice}
within knowledge itself. This character is expressed in the path
from the {\it perspective}, which lies in the {\it indeterminate
level}, to the {\it context}, which lies in the {\it determined
level}.\footnote{We must remark that in the quantum context there
is still certain indetermination regarding the properties in the
sense that a superposition expresses still the potential, and in
this sense {\it is} and {\it is not}.} Concepts are lacking in the
development of these levels.\footnote{For this purpose I have
introduced earlier (\cite{deRondeOP} and \cite{deRondeCDII}) the
concept of {\it ontological potentiality} as an attempt to
accompany quantum mechanics in its trip outside the limits of
classical thought.} The problem remains to provide a consistent
image (an \emph{anschaulich} content), of quantum theory. The
theory itself forces us to create new concepts which give a proper
account of what is quantum mechanics talking about. What does one
mean when one talks about a quantum wave function? What does one
mean when one talks about a quantum superposition?

\section{Paradoxes in Quantum Mechanics: Modality and Contextuality}

Quantum mechanics was born from a \emph{para-doxa}, which even
today remains still not completely recognized: the impossibility
of thinking about entities through the quantum formalism.

{\smallroman
\begin{quotation}
``[...] Mr. Bohr, is, in my opinion, the only truly non-Platonic
thinker: [footnote: The English philosopher {\it A. N. Whitehead}
(1861-1947) once said that the whole of European philosophy
consisted on footnotes to Plato] even in the early '20s (before
the establishment of present-day wave mechanics) he demonstrated
to me the pair of opposites ``Clarity-Truth" and taught me that
every true philosophy must actually start off with a {\it
paradox}. He was and is (unlike Plato) a {\it dekranos} [footnote:
``Double headed"--nickname for disciples of Heraclitus given by
disciples of Parmenides.] {\it kat exochen}, a master of antinomic
thinking.

As a physicist familiar with the course of development and this
way of thinking, the concepts of gentlemen with the stationary
spheres [footnote: I have {\it Parmenides} and {\it Kepler} in
mind.] are just as suspect to me as the concepts of ``being"
metaphysical spaces or ``heavens" (be they Christian or Platonic),
and ``the Supreme" or ``Absolute." [footnote: This is an allusion
to Indian philosophy. Even those Indian philosophers who, like
{\it Prof. S Radhakrishnan} (1888-1975), avoid applying the word
``illusion" to the empirical world have no other way of commenting
on the {\it Mysterium} of the connection between ``ultimate
reality" and the empirical world, except to call it ``Maya"] With
all these entities, there is an essential paradox of human
cognition (subject-object relation), which is not expressed, but
sooner or later, when the authors least expect it, it will come to
light!" W. Pauli [letter to C. G. Jung dated 27 February 1953]
(\cite{PauliJung}, p.93-94)
\end{quotation}}

In this section we want to analyze in more detail the constraints
which the formalism of quantum mechanics forces to accept, if we
are willing to talk about entities. The problem \emph{is not} how
can quantum mechanics be thought in terms of entities, but rather,
the problem is to find \emph{which are the conditions under which
quantum mechanics can be thought}.

\subsection{Modality in Quantum Mechanics}

The discussions regarding the interpretation of quantum mechanics
have not stopped since its birth in the year 1900. A way of
thinking about quantum mechanics is in terms of modalities, Max
Born interpreted the quantum wave function as expressing the
\emph{possibility} for a certain outcome to take place. In his
1926 article he writes:

{\smallroman
\begin{quotation}
``Schr\"odinger's quantum mechanics [therefore] gives quite a
definite answer to the question of the effect of the collision;
but there is no question of any causal description. One gets no
answer to the question, `what is the state after the collision'
but only to the question, `how probable is a specified outcome of
the collision'." M. Born (quoted from \cite{WZ}, p.57)
\end{quotation}}

\noindent Also Werner Heisenberg expressed the idea that the wave
function was something related to possibility but still not
completely clear:

{\smallroman
\begin{quotation}
``[...] the paper of Bohr, Kramers and Slater revealed one
essential feature of the correct interpretation of quantum theory.
This concept of the probability wave was something entirely new in
theoretical physics since Newton. Probability in mathematics or in
statistical mechanics means a statement about our degree of
knowledge of the actual situation. In throwing dice we do not know
the fine details of the motion of our hands which determine the
fall of the dice and therefore we say that the probability for
throwing a special number is just one in six. The probability wave
function of Bohr, Kramers and Slater, however, meant more than
that; it meant a tendency for something. It was a quantitative
version of the old concept of `potentia' in Aristotelian
philosophy. It introduced something standing in the middle between
the idea of an event ant the actual event, a strange kind of
physical reality just in the middle between possibility and
reality.''  W. Heisenberg (\cite{Heisenberg58}, p.42)
\end{quotation}}

It is interesting to characterize two main groups of
interpretations within the foundational geography of the quantum:
the first group presupposes certain ontology and tries to ``fit"
the formalism into their own metaphysical scheme (MW, Bohmian
mechanics, GRW, etc.). The second group does the opposite, namely,
it is interested in developing an interpretation which fits the
formalism. Modal interpretations pertain to this second group,
trying to learn about quantum mechanics, its structure and
meaning. In the following section, within the framework of the
modal interpretation, we would like to focus on the problem of
pasting together a holistic theory, such as quantum mechanics,
with a reductionistic theory, such as classical mechanics.

Contemporary modal interpretations have continued the footprints
left by Niels Bohr, Werner Heisenberg, Wolfgang Pauli and Max Born
and continued the path on the lines drawn by Bas van Fraassen,
Simon Kochen, Dennis Dieks and many others, searching for the
different possibilities of interpreting the formalism of the
theory.\footnote{See for example \cite{VF81, Kochen85, Dieks88}.}
Modal interpretations may be thought to be a study of the
constraints under which one is able to talk a consistent classical
discourse without contradiction with the quantum formalism.
Following the general characterization provided in
\cite{DFRStudies} one might state in general terms that a modal
interpretation is best characterized by the following points:

\begin{enumerate}
\item
One of the most significant features of modal interpretations is
that they stay close to the standard formulation. Following van
Fraassen's recommendation, one needs to learn from the formal
structure of the theory in order to develop an interpretation.
This is different from many attempts which presuppose an ontology
and then try to fit it into the formalism.

\item
Modal interpretations are non-collapse interpretations. The
evolution is always given by the Schr\"odinger equation of motion
and the collapse of the wave function is nothing but the path from
the possible to the actual, it is not considered a physical
process.

\item
Modal interpretations ascribe possible properties to quantum
systems. The property ascription depends on the states of the
systems and applies regardless of whether or not measurements are
performed.  There is a distinction between the level of
possibility and that of actuality which are related through an
interpretational rule.

\item
Modality is not interpreted in terms of ignorance. There is no
ignorance interpretation of the probability distribution assigned
to the physical properties. The state of the system determines all
there is to know. For modal interpretations there is no such thing
as `hidden variables' from which we could get more information.
One can formulate a KS theorem for modalities which expresses the
irreducible contextual character of the theory even in the case of
enriching its language with a modal operator.
\end{enumerate}

It is important to stress at this point that in the modal
interpretations one determines the set of definite properties
without adding anything by hand. Given the complete system and its
corresponding Hilbert space, the choice of what is the system
under study and what is the apparatus determines the factorization
of the complete space that leads to the set of definite properties
given by the Schmidt decomposition. In spite of the fact that this
cut is (mathematically) not fixed a priori in the formalism, the
(physical) choice of the apparatus determines explicitly the
context \cite{DFRStudies}. It is exactly this possibility, of
having different incompatible contexts given by the choice of
mutually incompatible apparatuses, which in turn determines KS
type contradictions within the modal interpretation (see for
discussion \cite{deRonde03}; and also \cite{Karakostas04}, section
6.1).

\subsection{Contextuality in the Modal Interpretation}

As expressed by van Fraassen: ``The most striking feature of
quantum theory is perhaps its {\it holism}: when a system is
complex, the state of the parts do not determine what the state of
the whole will be."\footnote{Quoted from \cite{vFraassen91},
p.73.} An expression of the holistic structure of the theory is
its contextual character which can be understood through the
Kochen Specker (KS) theorem \cite{KS}. Let's see this more in
detail. In quantum mechanics the wave function $\Psi$ is an
abstract mathematical form which can be expressed in different
{\it representations}, each of which is given in the formalism by
different basis $\{B, B', B'', ...\}$; each basis is conceived
here in the context of the modal interpretation, thus, providing
the set of properties which are determined. For each
representation we obtain respectively $\{|\Psi_{B}\rangle,
|\Psi_{B'}\rangle, |\Psi_{B''}\rangle,...\}$.\footnote{More
generally one can think in terms of density operators: firstly a
$\rho$ without a definite basis, and secondly, $\{\rho_{B},
\rho_{B'}, \rho_{B''},...\}$ given by the density operator in each
basis $\{B, B', B'', ...\}$.} We have to choose in which basis we
are going to write the wave function (context) just like in
classical mechanics we choose a certain reference frame to write
our equations of motion. But in quantum mechanics, contrary to
classical mechanics, each {\it representation/basis} expresses a
{\it context} which can be, in principle, \emph{incompatible} to a
different context. This is where all the trouble starts:
\emph{compatibility}.\footnote{For an analysis of the concept of
compatibility see for example the very interesting passage of the
book of Asher Peres (\cite{Peres93}, chapter 7.); see also
\cite{AertsdeRondeBart} and references therein.} Simon Kochen and
Ernst Specker  proved that in a Hilbert space $d\geq3$, it is
impossible to associate numerical values, 1 or 0, with every
projection operator $P_{m}$, in such a way that, if a set of {it
commuting} $P_{m}$ satisfies $\sum P_{m}=\amalg$, the
corresponding values, namely ${\it v}(P_{m})=0$ or 1, also satisfy
$\sum {\it v}(P_{m})=1$. This means that if we have three
operators $A$, $B$ and $C$, where $[A,B] = 0$, $[A,C] = 0$ but
$[B,C] \neq 0$ it is not the same to measure $A$ alone, or $A$
together with $B$, or together with $C$. In algebraic terms one
can state the KS theorem as follows \cite{DF05}:

\begin{theo}\label{CS3}
If $\mathcal{H}$ is a Hilbert space such that $dim({\cal H}) > 2$,
then a global valuation, i.e. a family of compatible valuations of
the contexts, over ${\mathcal L}({\mathcal H})$ is not possible.
\end{theo}

However, it is interesting to notice that KS theorem talks about
actual values of properties, while quantum mechanics seems to be
talking about modalities. The question we have posed, together
with Graciela Domenech and Hector Freytes is the following: Does
KS theorem have anything to say about possible values of
properties? Trying to answer this question a modal Kochen Specker
(MKS) theorem was developed which proves that contextuality can
not be escaped, even in the case modal propositions are taken into
account in the discourse \cite{DFRAnnalen}. The MKS theorem can be
stated as follows:

\begin{theo}\label{ksm}
Let $\cal L$ be an orthomodular lattice. Then $\cal L$ admits a
global valuation iff for each possibility space there exists a
Boolean homomorphism  $f: \diamond {\cal L} \rightarrow {\bf 2}$
that admits  a compatible actualization.
\end{theo}

In classical mechanics one talks with propositions which refer to
actual properties of a system, KS precludes this possibility in
quantum mechanics. On the other hand, classical statistical
mechanics refers to possible entities through an \emph{ignorance
interpretation} of modality, it is this possibility which is
untenable in quantum mechanics because of the MKS theorem.
\emph{It is not possible to say that one is talking in quantum
mechanics about possibility in terms of ignorance}. In our terms,
KS refers to the valuation of {\it actual} properties which
pertain to different contexts, while the MKS theorem refers to the
valuation of {\it possible} properties which pertain to different
contexts. The conclusion we believe must be drawn in general terms
from the MKS theorem is that the $\Psi$ cannot be thought in terms
of possible reductionistic contexts.\\

It is the holistic structure of quantum mechanics which is
responsible for the incompatibility of contexts; i.e. the
impossibility of assigning a compatible family of truth valuations
to the projection operators of different contexts, which brings
into stage the concept of {\it choice} (at least in relation to
the entity). If one wants to talk about entities in quantum
theory, subjectivity appears as a major problem. In quantum theory
the entity exists only because we choose. But this subjective
entity is unacceptable in science, which deals with objective
statements. In classical mechanics, on the contrary, due to its
compatible\footnote{Even though one might have incompatible
experimental setups (contexts) in classical mechanics, such as
those proposed by Aerts: A piece of wood which has the property of
being burnable and of floating \cite{Aerts81}. One can always
think of these contexts in terms of ignorance, there is no
proper/ontological incompatibility, as it is always possible in
principle to valuate every property without inconsistencies. It is
possible to think that the piece would {\it definitely has} the
mentioned properties.} (reductionistic) structure, one can neglect
this level (which we have called earlier ``perspective").
Reductionistic theories do not suffer from this ``problem" because
their structure always allows for a Boolean valuation. Coloring
every atom in the universe (every point in phase space) would not
arise a problem because the universe is nothing but the sum of
these atoms.\footnote{As it has been proved by several theorems
this is not possible to do in the quantum structure, see for
example K. Svozil demonstration in terms of a jigsaw puzzle
(\cite{Svozil99}, section 6).} Classically, the choice of the
context {\it discovers} (rather than {\it creates}) an element of
physical reality, which of course was already there... just like
the moon is outside there regardless of choice, of us looking at
her or not.\footnote{This is what Pauli used to call the
detachness of the observer, an ideal which Einstein wanted to
sustain but finds major difficulties in quantum mechanics.}

Within the modal interpretation proposed by Gyula Bene and Dennis
Dieks \cite{BeneDieks02} we have defined in \cite{deRonde03}, the
perspective, which describes all the {\it possible} (mutually
exclusive) contexts. When we choose a definite context we obtain a
definite relation between subsystems and the (Boolean) classical
structure is {\it almost} regained. It is still not completely
regained as the contexts remain fundamentally holistic. The
properties arising from them are best characterized by an improper
mixture, as we have seen through the MKS theorem, it is not
possible to interpret the contexts as reductionistic contexts. One
still needs to make an ``interpretational jump" and forget about
the procedure through which one obtained this mixture. Only then
one is able to give an ignorance interpretation and talk {\it as
if} these properties where reductionistic. We then describe
everything {\it as if} what we had obtained was an proper mixture
(which later describes an entity of which we are uncertain).

In the complementary descriptions approach we have argued that
these descriptions are mutually objective in the sense just
explained and only together give a deeper understanding of the
correlations in quantum mechanics. As mentioned earlier we have
introduced the concept of context as the definite factorization
while the perspective remains as the condition of possibility for
this choice to take place (\cite{deRonde03}, section 7.1). These
levels are necessary not only as a pedagogical source but as the
determination of a new conceptual scheme which in turn produces a
new expression of reality. It is through this development that we
have analyzed the measurement process in quantum mechanics (see
\cite{deRondeCDII}, section 2). The determination of the context
is the final ``Heisenberg cut" which determines the transition
from the perspective level to the context level:

{\smallroman
\begin{quotation}
``This is indeed the actual situation [in quantum mechanics]
created by the finiteness of the quantum of action. One is here,
as {\it Heisenberg} first pointed out, always in the position of a
dilemma between the sacrifice and the choice, a situation which
implies a certain freedom on the side of the observer to choose
his experimental arrangement as one of at least two possibilities,
excluding each other." W. Pauli (\cite{Pauli94}, p.32)
\end{quotation}}

The perspective can be factorized in infinitely many ways, each of
which determines a definite relation between all the subsystems. A
complete set of `new' contexts appear each time we choose to
change the factorization. The perspective cannot be {\it a priori}
decomposed into elementary blocks, these holistic contexts, and
the whole from which they `become', should be regarded as
expressing the essential character of quantum mechanics, that of
precluding the possibility of thinking about the quantum wave
function in terms of the classical principles of identity, unity
and totality.

\subsection{Green Tables and Boeings 747 in the Modal Interpretation}

Property Composition ($PC$) and Property Decomposition ($PD$) are
common features of our everyday reasoning and intuitions (see
\cite{Clifton96}, p.385); these properties follow in classical
Boolean logic and we take them for granted since first grade
school when we are introduced with `Venn diagrams' and `set
theory'.\footnote{\emph{Property Composition}: given a vector $P
\in S$, then $P\otimes I \in S+S'$ and for all truth valuations
$[P \otimes I] = [P]$. \emph{Property Decomposition}: given a
vector $P\otimes I \in S+S'$, then $P \in S$ and for all truth
valuations $[P] = [P \otimes I].$} However, these properties do
not follow in every kind of logic, more specifically, they do not
follow in quantum (non-Boolean) logic. In this kind of logic,
firstly introduced by Birkhoff and J. von Neumann \cite{BvN}, the
Distributive Condition ($DC$) is not fulfilled.\footnote{The
distributive condition states that: $A$ and $(B$ or $C)= (A$ and
$B)$ or $(A$ and $C)$.} Quantum properties do not follow classical
Boolean logic, but a non-distributive logic which corresponds to
the algebra of closed subspaces of vector Hilbert space, with
`meet' and `joint' operations corresponding to `intersection' and
`direct sum' of subspaces. What one calls ``electron" in quantum
mechanics does not behave like any kind of entity. The problem is
that our language leaves little space for anything which is not an
entity, so we have no other choice than to presuppose from the
very beginning that an electron is an entity. In close relation to
Einstein's idea that separability should be regarded as a
presupposition in doing physics, Arntzenius \cite{Arntzenius90,
Arntzenius98} and Clifton \cite{Clifton95c, Clifton96} have argued
that the absence of $PC$ and $PD$ in an interpretation of quantum
mechanics (more specifically in the K-D and D-V modal
interpretations) makes it {\it metaphysically untenable}. Contrary
to this idea, we have stressed that one should not take as
presuppositions, the conditions under which quantum mechanics can
provide a general picture, rather, one should stay close to the
quantum formalism and find out what quantum mechanics is trying to
tell us. Although these conditions might seem ``common sense"
conditions, it is exactly this common sense which quantum
mechanics seems to go against.

Arntzenius \cite{Arntzenius90} discusses the example of assigning
properties to a green table in which $Q_{a}$ represents the
property `greenness' of the left hand side of the table, and
$Q_{a}I$ represents the property `greenness' of the table as a
whole. The thing is that these properties $Q_{a}$ and $Q_{a}I$
meet the demands of being different and of being observationally
indistinguishable: from the logical point of view $Q_{a}$
represents the proposition `The left hand side of the table is
green' and $Q_{a}I$ represents the proposition `The table as a
whole is green at the left hand side'.\footnote{It is interesting
to notice the fact that the language already tricks the example as
there is no word for ``half a table"; Bohr's dictum finds here a
clear example of the fact that {\it ``we are suspended in
language"}.} These propositions can be analyzed as predicating two
different predicates, namely, `green' and `green at the left hand
side', respectively, to two different individuals, `the left hand
side of the table' and `the table as a whole', respectively. By
means of this example, Arntzenius discusses the failure of $PC$
and argues that the fact that different truth values are assigned
to propositions like `the left hand side of the table is green'
and `the table has a green left side' is bizarre.

Clifton (\cite{Clifton96}, section 2.3) developed an example in
which the violations of $PC$ and $PD$ seem to show implications
which seem at least incompatible with the everyday description of
reality. In the example Clifton takes a Boeing 747 which has a
possibly wrapped left-hand wing: $a$ is the left-hand wing and
$a_{\beta}$ is the airplane as a whole. $Q_{a}$ represents the
property of `being wrapped' and $Q_{a}I_{\beta}$ represents the
`plane property of the left wing being wrapped'. In such an
example a violation of $PC$ ($[Q_{a}]=1$ and $[Q_{a}I_{\beta}]\neq
1$) leads, according to Clifton (\cite{Clifton96}, p.385.), to the
following situation: {\it ``a pilot could still be confident
flying in the 747 despite the fault in the left hand wing"}. If,
on the other hand, $PD$ fails ($[Q_{a}]\neq1$ and
$[Q_{a}I_{\beta}]=1$) the implication reads {\it ``no one would
fly in the 747; but, then again, a mechanic would be hard-pressed
to locate any flow in its left-hand wing"}. The situation gets
even stranger when the pilot notices that the plane as a whole has
the property $[Q_{a}I_{\beta}]=1$ and concludes (incorrectly)
following $PD$ that the left-hand wing is wrapped, that is, that
$[Q_{a}]=1$. The mechanic is then sent to fix the left hand-side
wing but according to Clifton cannot locate the flaw because the
wing does not possess the property $Qa$. Arntzenius explains the
situation as follows:

{\smallroman
\begin{quotation}
``One should view the mechanic as having a list of all the
definite properties on the left hand wing handed to him, e.g. by
God, while the pilot says to the mechanic: `Hmm, the left hand
wing is wrapped, that's a problem'. The mechanic responds: `No,
I've got all the properties of the left-hand wing, and nowhere is
it listed that it is wrapped'. This seems bizarre. [...] It does
not appear to get any less bizarre if Vermaas, standing next to
the pilot and mechanic offers the following advice: `But the two
of you are talking about different systems. If you are careful
about this you will find that what you are saying is
inconsistent'. True, if all claims have to be system-dependent in
this manner the claims are consistent, but it remains a bizarre
system-dependent world of properties. [...] One may get into the
situation where it's true of the left wing that it is dropped of
the plane, while it is not true of the entire plain that it has
lost its left wing. That would be strange..." Arntzenius
(\cite{Arntzenius98}, p.370)
\end{quotation}}

For us, the problem remains how to think quantum mechanics
consistently. Rather than adding conditions of ``common sense" we
should wonder about how to understand the conditions which quantum
mechanics already respects. We cannot expect that fundamentally
new theories (like we think is the case of quantum mechanics)
respect the logical principles of classical physics; this would be
tantamount to demand that classical physics should determine the
conceptual tools of new theories. As noted by Dieks, this would
deny the possibility of really new fundamental theories,
conceptually independent of classical physics. Riemann geometry
goes against ``common sense", in this geometry the idea that two
parallels do not intersect is left aside. Earlier mathematicians
were looking for an {\it ad absurdum} demonstration of the
impossibility to leave the fifth Euclidean axiom. Riemann built a
new geometrical system as consistent as its predecessor, however,
no one would say today that Non-Euclidean geometry is ``nonsense",
least of all physicists, whom have used this geometry to develop
relativity theory, one of the most important theories created in
the last century. Analogously, we believe that a deeper
understanding of the logic inherent to quantum mechanics can help
us in providing an image of the theory.

Our answer is straightforward, an airplane is a classical concept
in itself which presupposes from the very beginning our classical
conception of the world. The tacit idea surrounding this
discussion is reductionism. Quantum mechanics should not
necessarily be regarded as a theory which refers to airplanes, or
in other words, there is no need that every theory is mapped into
classical physics. The 747 will have reductionistic properties,
the mechanic and the pilot must agree, simply because they are
talking about an airplane which presupposes a classical
description. It is only the theory which can tell you what can be
measured, and the presuppositions for the existence of an airplane
lie within the walls of classicality. The mistake is to believe
that an airplane is something which exists independently of the
description. An airplane is an expression of classical physics and
thus, follows its logic. Quantum mechanics can not be applied to
an airplane, this is in no way different from the fact that
quantum mechanics leads to inconsistencies when being applied to a
classical apparatus. Everything that we have learned until today
points directly to the idea that quantum mechanics cannot be
subsumed into classical thought.

\subsection{Ontological Potentiality in the Modal Interpretation}

The meaning of {\it possibility} has been a matter of debate
within the modal interpretation of quantum mechanics. For example,
Bacciagaluppi has stated the following: ``I would claim that,
despite the name, the modal interpretation in the version of
Vermaas and Dieks is a theory about actualities -- albeit a
stochastic one." Van Fraasen, even though distinguishes between a
dynamical state and a value state, takes modalities to be only
semantical. Regarding van Fraassen: \emph{There is only one actual
reality. Modalities are in our theories, not in the world}.
Contrary to these views, within the complementary descriptions
framework we regard the quantum description as describing the
possible from an equal ontological standpoint to the classical
description. In this sense the formalism of quantum mechanics, and
therefore a superposition state reflects a character of reality
just in the same way a Stern-Gerlach apparatus does; as discussed
in section 2.3 there is no actuality voided of description as
theoretical presuppositions. It is important to notice that modal
interpretations, as they stand, do not talk about 'Schr\"odinger
cats', superpositions, as ontologically existent entities, they
only talk about actual cats (!) We wish to stand for the opposite:
\emph{Modalities are in the world, not only in our theories, they
express a complementary description of reality.} This determines
an empirical distinction between our modal interpretation of
quantum mechanics and those which regard superpositions as void of
an ontological status (see \cite{deRondeCDII}, section 1.4).

Through the complementary descriptions approach we have presented
earlier the idea that the problem posed by Vermaas:

{\smallroman
\begin{quotation}
``There does not exist a super modal interpretation which
integrates the holistic and reductionistic features of the bi
(K-D), spectral (D-V) and atomic modal interpretations. Instead
one has to choose and settle for either a holistic or
reductionistic interpretation" P. Vermaas (\cite{Vermaas99},
p.233) \end{quotation}}

\noindent can be solved by taking into account both, holistic and
reductionistic descriptions \cite{deRonde03, deRondeOP,
deRondeCDI, deRondeCDII}. The holistic context talks about the
potential in an ontological stance, as something which cannot be
reduced to the actual; while the reductionistic context talks
about the possible as related to the actual.

We believe that the impossibility of solving this problem appears
from regarding the actual as the real. Thinking the potential is
merely the possible, that it can be reducible to actuality. The
idea of regarding actuality as the real is a heavy burden in
western though which comes already from Aristotelian philosophy
and its cosmology. Even though Aristotle understood the problem of
movement, and created the concept of \emph{potential being} his
choice of grounding existence on the immobile motor, \emph{pure
acto}, determined the path of philosophy on the lines of actuality
(as opposed to potentiality).

{\smallroman
\begin{quotation}
``Aristotle [...] created the important concept of {\it potential
being} and applied it to {\it hyle}. [...] This is where an
important differentiation in scientific thinking came in.
Aristotle's further statements on matter cannot really be applied
in physics, and it seems to me that much of the confusion in
Aristotle steams from the fact that being by far the less able
thinker, he was completely overwhelmed by Plato. He was not able
to fully carry out his intention to grasp the \emph{potential},
and his endeavors became bogged down in early stages." W. Pauli
(\cite{PauliJung}, p.93)\end{quotation}}

The problem which we recover from Pauli and Heisenberg is this:
\emph{how is it possible to think in terms of potentiality?}
Aristotle distinguishes between two types of potentiality.
Firstly, {\it generic potentiality}: the potentiality of a seed
that can transform into a tree; this generical sense is not that
which interests Aristotle but rather the second possibility, the
{\it potentiality as a mode of existence}: the poet has the
capacity of writing poems but also of not writing poems. It is not
only the potentiality of doing this or that, but also the
potentiality of not-doing, potentiality of not being, of not
passing to the actual. What is potential is capable of being and
not being. This is the problem of potentiality: {\it the problem
of possessing a faculty.} What do I mean when I say ``I can", ``I
cannot". Ontological potentiality is a mode of existence which
expresses \emph{power to do}, \emph{pure action}.\footnote{Earlier
\cite{deRondeOP} we have used the ideas of Agamben
(\cite{Agamben99}, p.183) in order to read Aristotle: ``What is
essential is that potentiality is not simply non-Being, simple
privation, but rather the existence of non Being, the presence of
an absence; this is what we call a `faculty' or `power.' `To have
a faculty' means to have a privation. And potentiality is not a
logical hypostasis but the mode of existence of this privation.
But how can an absence be present, how can a sensation exist as
anesthesia? This is the problem that interests Aristotle."  We
would like now to take distance from the negative part of this
interpretation as we consider that ontological potentiality should
be thought in terms of \emph{power to do}, \emph{pure affirmation}
(see also \cite{GdeRPot}).} In order to develop the quantum
description we have introduced the concept of {\it ontological
potentiality} which presents us with, as Aristotle expressed, a
different form of the being. The central point of the concept of
{\it ontological potentiality} is that it cannot be reduced to
{\it actuality} and presents us with a different form of the
being, i.e. {\it the necessity of considering potentiality as
ontologically independent of actuality}. I think that in order to
reach a deep understanding of the quantum theory it is necessary
to go further and develop this concept which allows us to think
quantum mechanics in a new fresh way. Wolfgang Pauli had foreseen
this path and pointed directly to it:

{\smallroman
\begin{quotation}
``Science today has now, I believe, arrived at a stage were it can
proceed (albeit in a way as yet not at all clear) along the path
laid down by Aristotle. The complementarity characteristics of the
electron (and the atom) (wave and particle) are in fact
``potential being," but one of them is always ``actual nonbeing."
{\it That is why one can say that science, being no longer
classical, is for the first time a genuine theory of becoming and
no longer Platonic.}" W. Pauli (\cite{PauliJung},
p.93)\end{quotation}}

\section{Entities and Faculties in Quantum Mechanics}

Going back to the VI century B.C. we can see that the seed of
occidental physics was already present in the problem of movement.
As the discussion has come to our days, specially through the
reading of Plato and Aristotle,\footnote{One could state following
K. Verelst \cite{VerelstCoecke} that: ``the `contradiction' seen
by classical philosophy between Heraclitus and Parmenides is not
necessarily a correct understanding of the earlier `philosophies'.
One could as well infer that Heraclitus and Parmenides do
articulate the same world-experience, the former as the experience
of reality over a lapse of time, the latter as the experience of
the absolute reality of this moment."} the pre-Socratic thought
had approached this problem from two, seemingly opposed positions.
Hercalitus of Elea, stated the theory of flux, a doctrine of
permanent motion and unstability in the world. The consequences of
this doctrine are, as both Plato and Aristotle stressed
repeatedly, the impossibility to develop stable, certain knowledge
about the world, for an object, changing each instant, does not
allow for even to be named with certainty, let alone to be
`known', i.e., assigned fixed, objective characteristics.
Parmenides was placed at the opposite side, teaching the
non-existence of motion and change in reality, reality being
absolutely One, and being absolutely Being \cite{VerelstCoecke}.
Aristotle's solution to the problem of movement soon became the
guiding line of physics. It was this solution which determined the
fate of western thought. The idea of Aristotle was to postulate
the existence of \emph{essential properties} which determined the
essence of that which existed. An entity is thus, something which
has a distinct, separate existence, a collection of characters
which can be placed as a unity, an identity, a totality. Stability
of the being was placed in the entity, becoming was lost.
Following Verelst and Coecke:

{\smallroman
\begin{quotation}
``We are tempted to see the origins of the QM-paradoxes as
consequences of the ontological `choices' of Plato and Aristotle.
Their effort concerned the stabilization of the world of constant
change, thus saving the possibility of certain knowledge in order
to escape the contradictions between stable and unstable, knowable
and unknowable that appear on the level of what happens in
reality, as expressed mainly by Heraclitus and Parmenides." K.
Verelst and B. Coecke (\cite{VerelstCoecke}, p.165)
\end{quotation}}

In order to point directly to the idea of entity we must recognize
its fundament in the principles of classical (Aristotelian) logic:
the existence of objects of knowledge, the principle of
contradiction and the principle of identity which are exemplified
in the three possible usages of the verb ``to be": existential,
predicative, and identical.

{\smallroman
\begin{quotation}
``The Aristotelian syllogism always starts with the affirmation of
existence: something \emph{is}. The principle of contradiction
then concerns the way one can speak (predicate) validly about this
existing object, i.e., about the truth and falsehood of its having
properties, not about its being in existence. The principle of
identity states that the entity is identical to itself at any
moment (a=a), thus granting the stability necessary to name
(identify) it. It will be clear that the principle of
contradiction and the principle of identity are closely
interconnected. In any way, \emph{change and motion are
intrinsically not provided for in this framework; therefore the
ontology underlying the logical system of knowledge is essentially
static}, and requires the introduction of a First Mover with a
proper ontological status beyond the phenomena for whose change
and motion he must account." K. Verelst and B. Coecke
(\cite{VerelstCoecke}, p.172, emphasis added)
\end{quotation}}

It is the idea of entity which generated the development of
physics since Aristotle. This is why we might say today that the
history of classical physics is the history of physical entities:
particles, waves, fields, etc. Even though we recognize its
importance in occidental thought, we believe that the idea of
entity appears, in the context of quantum mechanics as, what
Gaston de Bachelard \cite{Bachelard} calls, an ``epistemological
obstruction"; i.e. an idea which restricts our possibilities to
imagine the physics provided by quantum mechanics.

This conception of Nature was taken to its lasts consequences in
the modern period which started in the XVII century. Nature itself
had become an entity which needed to be studied, ``cutted into
pieces". The Baconian principle of \emph{disectum naturae} became
the methodological weapon of modern times. As noted by Heisenberg
these ideas were a direct consequence of the reading of the
Cartesian partition as a cut between ``res cogitans" and ``res
extensa".

{\smallroman
\begin{quotation}
``If one uses the fundamental concepts of Descartes at all, it is
essential that God is in the world and in the I and it is also
essential that the I cannot be really separated from the world. Of
course Descartes knew the undisputable necessity of the
connection, but philosophy and natural science in the following
period developed on the basis of the polarity between the `res
cogitans' and the `res extensa'. The influence of the Cartesian
division on human thought in the following centuries can hardly be
overestimated, but it is just this division which we have to
criticize later from the development of physics in our time.

[...] Natural science does not simply describe and explain nature;
it is a part of the interplay between nature and ourselves; it
describes nature as exposed to our method of questioning. This was
a possibility of which Descartes could not have thought, but
\emph{it makes the sharp separation between the world and the I
impossible}.

If one follows the great difficulty which even eminent scientists
like Einstein had in understanding and accepting the Copenhagen
interpretation of quantum theory, one can trace the roots of this
difficulty to the Cartesian partition. This partition has
penetrated deeply into the human mind during the three centuries
following Descartes and it will take a long time for it to be
replaced by a really different attitude toward the problem of
reality.

The position to which the Cartesian partition has led with respect
to the `res extensa' was what one may call metaphysical realism.
The world, i.e. the extended things, `exist'. [...] Dogmatic
realism claims that there are no statements concerning the
material world that cannot be objectivated. [...] When Einstein
has criticized quantum theory he has done so from the basis of
dogmatic realism. This is a very natural attitude. Every scientist
who does research work feels that he is looking for something that
is objectively true. His statements are not meant to depend upon
the conditions under which they can be verified. Especially in
physics the fact that we can explain nature by simple mathematical
laws tells us that here we have met some genuine feature of
reality, not something which we have--in any menaing of the
world--invented ourselves. This is the situation which Einstein
had in mind when he took dogmatic realism as the basis for natural
science. But quantum theory is in itself an example for the
possibility of explaining nature by means of simple mathematical
laws without this basis." W. Heisenberg (\cite{Heisenberg58},
p.73)\end{quotation}}

However, though we see the critics of Heisenberg as touching an
important point, we also believe that Albert Einstein clearly
understood the radicalness of the quantum revolution regarding the
problem of physical reality. If one takes seriously the formalism
of quantum mechanics the conclusion that must be drawn is that
entities are some kind of subjective elements of physical reality.
The ontological concern which he repeatedly posed to Born,
Heisenberg, Pauli and specially to Bohr, did not find an adequate
answer. His concern, which is nowadays commonly charicaturized,
was absolutely justified.

{\smallroman
\begin{quotation}
``\emph{Einstein}'s opposition to [quantum mechanics] is again
reflected in his papers which he published, at first in
collaborations with \emph{Rosen} and \emph{Podolsky}, and later
alone, as a critique on the concept of reality in quantum
mechanics. We often discussed these questions together, and I
invariably profited very greatly even when I could not agree with
\emph{Einstein}'s view. ``Physics is after all the description of
reality" he said to me, continuing, with a sarcastic glance in my
direction ``or should I perhaps say physics is the description of
what one merely imagines?" This question clearly shows
\emph{Einstein}'s concern that the objective character of physics
might be lost through a theory of the type of quantum mechanics,
in that as a consequence of a wider conception of the objectivity
of an explanation of nature the difference between physical
reality and dream or hallucination might become blurred." W. Pauli
(\cite{Pauli94}, p.122)\end{quotation}}

The truth is that in quantum mechanics it is not possible to state
that $\Psi$ refers to some kind of entity which exist
independently of our choice (see \cite{Karakostas04} for a
detailed discussion). The quantum wave function $\Psi$ is a
non-represented element in which mutually incompatible
representations co-exist. In order to solve the paradox we claim
that quantum mechanics does not talk about \emph{entities}. It
might be thus regarded as the first physical theory which does not
refer objectively to some kind of entity (particles, waves,
fields, etc.). Quantum mechanics talks about \emph{faculties} in
an objective way \cite{deRondeFaculties}. The ontological problem
of understanding the meaning of the quantum wave function can be
only solved if we are able to create a new way of relating to
Nature. A way which does not avoid recognizing the limitations of
the classical view of the world.

We have introduced in \cite{deRondeCDI, deRondeCDII} the concept
of \textbf{faculty}, in order to give account of that of which
quantum mechanics is talking about. The mode of being of a faculty
is potentiality, not thought in terms of possibility (which relies
on actuality) but rather in terms of what we have called
ontological potentiality: something which \emph{is} and \emph{is
not}, \emph{here} and \emph{now}. A faculty is determined by a
certain state of affairs, and in this sense it is from the very
start a contextual concept. I have the faculty of {\it raising my
hand}, which does not mean that ``{\it I will} raise my hand" or
``{\it I will not} raise my hand"; what it means is that here and
now I possess a faculty which exists in the mode of being of
potentiality, independent of what will happen in actuality. A
faculty is determined always by a certain state of affairs. I
possess the faculties of {\it running} and {\it swimming}, but in
order for these faculties to exists, I must be either in a place
where either I can run or I can swim. I can say: ``\emph{I can}
swim (here and now)" only if I am in a place where I can swim,
like for example in a swimmingpool. This has nothing to do with
the fact that in the near future I choose either to swim or not to
swim while I'm in the swimmingpool. In a swimmingpool however, I
am not able to run, just in the same way that I am not able to
swim in the street. In our earlier terms the context determines
the existence of the faculty explicitly, faculties are contextual
existents. We can also see from this example tat {\it
incompatibility} is a central feature of faculties. Our strong
claim is that faculties do exist in quantum physics just like
entities exist in the realm of classical physics. An entity is
governed by the logical principles of classical logic (principle
of existence, principle of non-contradiction and principle of
identity) but a faculty is governed by completely different
principles, namely, Heisenberg's principle of indetermination,
Bohr's principle of
complementarity and the superposition principle.\\

We are now in conditions to complete our schematic
approximation:\\

\begin{tabular}{|c|c|c|c|c|}

\hline
& \tiny{\textbf{PERSPECTIVE}} & \textbf{\tiny{HOLISTIC}} & \tiny{\textbf{REDUCTIONISTIC}} & \tiny{\textbf{MEASUREMENT}} \\
&  & \textbf{\tiny{CONTEXT}} & \tiny{\textbf{CONTEXT}} & \tiny{\textbf{RESULT}} \\
\hline
\tiny{\textsl{MODE OF BEING}} & \footnotesize{{\it potential}} & \footnotesize{{\it potential}}  & \footnotesize{{\it possible/probable}} & \footnotesize{{\it actual}} \\
\hline
\tiny{\textsl{FORMAL EXPRESSION}} & $\Psi$ & $\psi_{B}$ & $\psi_{B}$ & $\alpha_{k}$, $|\alpha_{k}\rangle$ \\
\hline
\tiny{\textsl{THEORETICAL EXPRESSION}} & \footnotesize{{\it potential mixtures}} & \footnotesize{{\it improper mixture}} & \footnotesize{{\it proper mixture}} & \footnotesize{{\it single term}}\\
\hline
\tiny{\textsl{PROPERTY}} & -- & \footnotesize{{\it holisic/non-Boolean/}} & \footnotesize{{\it reductionistic/Boolean/}} & \footnotesize{{\it actual}} \\
 &  & \footnotesize{{\it superposition}} & \footnotesize{{\it ensemble}} &  \\
\hline
\tiny{\textsl{DESCRIPTION IN TERMS OF}} & \footnotesize{{\it potential faculties}} & \footnotesize{{\it faculty}} & \footnotesize{{\it possible entity}} & \footnotesize{{\it actual entity}} \\
\hline
 & \footnotesize{{\it indetermination}} & \footnotesize{{\it indetermination}} & \footnotesize{{\it existence}} & \footnotesize{{\it existence}} \\
\tiny{\textsl{LOGICAL PRINCIPLES}} & \footnotesize{{\it complementarity}} & \footnotesize{{\it complementarity}} & \footnotesize{{\it non-contradiction}} & \footnotesize{{\it non-contradiction}} \\
 & \footnotesize{{\it superposition}} & \footnotesize{{\it superposition}} & \footnotesize{{\it identity}} & \footnotesize{{\it identity}} \\
\hline
\end{tabular}

\section{Discussion}

According to the received view our language presupposes
individuality, it is an expression of the metaphysical choice
inherited from Aristotle, Plato and the Parmenidean `One'. These
presuppositions expressed by classical logic go together with
classical mechanics. Quantum mechanics, on the other hand, is
closer to the Heraclitean `Many'.\footnote{See for example David
Finkelstein's paper: ``All is Flux" for a related analysis
\cite{Finkelstein87}.} Classical and quantum mechanics are
different expressions of Nature. Both theories approach reality,
the result of which is to a large extent determined by the
description itself: they are mutually incompatible, and at the
same time they are expressions of the preconceptions involved
presenting complementary views of reality. It is in this general
sense that I have expressed the idea that {\it the quantum
mechanical description can be regarded as complementary to the
classical description}.

As expressed by Folse (\cite{Folse87}, p.163): ``Bohr persistently
evades any direct engagement with the question of `reality'." This
is why, in this specific sense, we are closer to Einstein's
concern about ontology. Physics deals with the questions about
what there \emph{is} and not only about what we can merely
\emph{say}. Off course we do not neglect the importance of the
linguistic turn, but we do not regard language as being all there
is. Physics deals with the problem of reality, but reality should
not be a pre-established concept nor a prejudice in observing and
relating empirical data, rather it should remain a goal concept
which should be transformed and developed. We should not expect
reality to be... as we would like it to be; we must constantly
revise the conceptual framework with which such a description is
expressed. Following the main idea, which led Einstein to the
special theory of relativity, we should not conclude experiments
from reality; however, the opposite should be neither pursued.
Carlo Rovelli (\cite{Rovelli96}, p.2) proposed the following: ``I
have a methodological suggestion for the problem of the
interpretation of quantum mechanics: Finding the set of physical
facts from which the quantum mechanics's formalism can be
derived." The problem is: there are no {\it facts} without a {\it
description}. As Einstein himself pointed out to Heisenberg: ``It
is only the theory which decides what can be observed."
Experimental observation and description are intricately related
without supremacy of one over the other, both being the reflection
of the former like two mirrors with nothing in between.

My standpoint is that experience is defined by description, {\it
and vice versa}, description is defined by experience, they
intricate themselves with no preponderance of the one over the
other. In order to regain objectivity, we must acknowledge that
the classical description has no supremacy over different
descriptions, that it is not given to us {\it a priori}, and that
it develops through the different descriptions with which we
choose to express ourselves. The phrase of Bohr (quoted from
\cite{WZ}, p.7) stating that ``the unambiguous interpretation of
any measurement must be essentially framed in terms of the
classical physical theories, and we may say that in this sense the
language of Newton and Maxwell will remain the language of physics
for all time." gives way to a quite strong conclusion, this is,
that we have come to the limits in the pre-conditions of human
understanding. By having to confront Einstein's realism and
ontological position Bohr was dragged to the other extreme,
namely, a purely epistemological position. We hope to go back to
the {\it middle path}, just in between experience and description,
between creation and discovery.\footnote{A related position,
namely, the creation discovery view was introduced by Aerts in
\cite{Aerts88}.}

We claim that the distinction between perspective and context in
order to describe the quantum measurement problem and the path
from the \emph{potential} to the \emph{actual} (see
\cite{deRondeCDII}, section 1) is central in order to understand
the path from the quantum description to the classical description
and {\it vice versa}. Our interpretation brings into stage a new
reading of how to understand modality in quantum mechanics. In
this sense, with our concept of ontological potentiality, we take
distance from earlier conceptions of potentiality in terms of
\emph{becoming actual} (which is the first distinction of
Aristotle) such as those proposed by Heisenberg, Piron, Popper,
Margenau and more recently by Aerts and Su\'arez (see also
\cite{Karakostas04, Smets05} for further discussions on the
subject). This concepts use only the first type of potentiality
which is that which needs actuality as a presupposition.
Potentiality should be understood as a mode of the being, but in
order to escape its interpretation in terms of actuality the idea
of \emph{faculty} becomes fundamental. Finally, our approach
discusses the concept of faculty as expressing that of which
quantum mechanics is talking about. The articulation of faculties
within the scope of quantum mechanics will be discussed in more
detail in \cite{deRondeFaculties}.

As Pauli stated, we must confront the most important problem of
our time, which is the problem of developing a new idea of
reality. Only through a new idea of reality is that we will be
able to take into account the new experience to which quantum
mechanics confronts us. Only through this development is that we
can think a fresh new experiments which can lead to new
discoveries. Only through these developments is that we can learn
to understand quantum mechanics.

\section*{Acknowledgements}

I wish to thank Graciela Domenech, Fernando Gallego, Federico
Holik and Karin Verelst for the many discussions and comments on
earlier drafts of this article and related topics. I also wish to
thank Belen Asad for a careful reading of an earlier draft. The
research for this paper benefited from the Projects of the Fund
for Scientific Research Flanders G.0362.03 and G.0452.04.

\end{document}